\documentclass[10pt,sigconf,nonacm]{acmart}
\PassOptionsToPackage{dvipsnames}{xcolor}

\usepackage{graphicx}
\usepackage{balance}  % for  \balance command ON LAST PAGE  (only there!)
\usepackage{booktabs} % For formal tables
\usepackage{balance}  % for  \balance command ON LAST PAGE  (only there!)
\usepackage{subcaption}
\usepackage{enumitem}
\usepackage{mdwlist}
\usepackage{hyperref} % for \url
\usepackage{cleveref}
\usepackage{cuted}

\setcopyright{none}

\newcommand{\system}{Chiller}
\newcommand{\nowait}{\texttt{NO\_WAIT}}
\newcommand{\waitdie}{\texttt{WAIT\_DIE}}
\newcommand{\occ}{\texttt{OCC}}

\usepackage[font=small]{caption}

\usepackage[compact]{titlesec}
\titlespacing*{\section}{0pt}{2pt}{2pt}
\titlespacing*{\subsection}{0pt}{1pt}{1pt}
\titlespacing*{\subsubsection}{0pt}{1pt}{1pt}

\usepackage{enumitem}
\usepackage{microtype}

\begin{document}
\fancyhead{} % do not delete this code.

\title{Chiller: Contention-centric Transaction Execution and Data Partitioning for Modern Networks}

\author{Erfan Zamanian}
\email{erfanz@cs.brown.edu}
\affiliation{%
  \institution{Brown University}
}

\author{Julian Shun}
\email{jshun@mit.edu}
\affiliation{%
  \institution{MIT CSAIL}
}

\author{Carsten Binnig}
\email{cbinnig@cs.tu-darmstadt.de}
\affiliation{%
  \institution{TU Darmstadt}
}

\author{Tim Kraska}
\email{kraska@csail.mit.edu}
\affiliation{%
  \institution{MIT CSAIL}
}

\begin{abstract}
Distributed transactions on high-overhead TCP/IP-based networks were conventionally considered to be prohibitively expensive and thus were avoided at all costs.
To that end, the primary goal of almost any existing partitioning scheme is to minimize the number of cross-partition transactions.
However, with the new generation of fast RDMA-enabled networks, this assumption is no longer valid.
In fact, recent work has shown that distributed databases can scale even when the majority of transactions are cross-partition.

In this paper, we first make the case that the new bottleneck which hinders truly scalable transaction processing in modern RDMA-enabled databases is \textit{data contention}, and that optimizing for data contention leads to different partitioning layouts than optimizing for the number of distributed transactions. 
We then present Chiller, a new approach to data partitioning and transaction execution, which aims to minimize data contention for both local and distributed transactions. 
Finally, we evaluate Chiller using various workloads, and show that our partitioning and execution strategy outperforms traditional partitioning techniques which try to avoid distributed transactions, by up to a factor of 2.
\end{abstract}

\maketitle

\section{Introduction}
\label{sec:intro}

The common wisdom is to avoid distributed transactions at almost all costs as they represent the dominating bottleneck in distributed database systems.
As a result, many partitioning schemes have been proposed with the goal of minimizing the number of cross-partition transactions~\cite{curino2010schism,tran2014jecb, pavlo2012skew, serafini2016clay,taft2014store, zamanian2015locality}.
Yet, a recent result~\cite{zamanian2017namdb} has shown that with the advances of high-bandwidth RDMA-enabled networks, neither the message overhead nor the network bandwidth are limiting factors anymore, significantly mitigating the scalability issues of traditional systems. 
This raises the fundamental question of how data should be partitioned across machines given high-bandwidth low-latency networks. 
In this paper, we argue that the new optimization goal should be to minimize contention rather than distributed transactions.

\begin{figure*}
    \centering
    \begin{minipage}{0.43\textwidth}
        \centering
        \includegraphics[width=\textwidth]{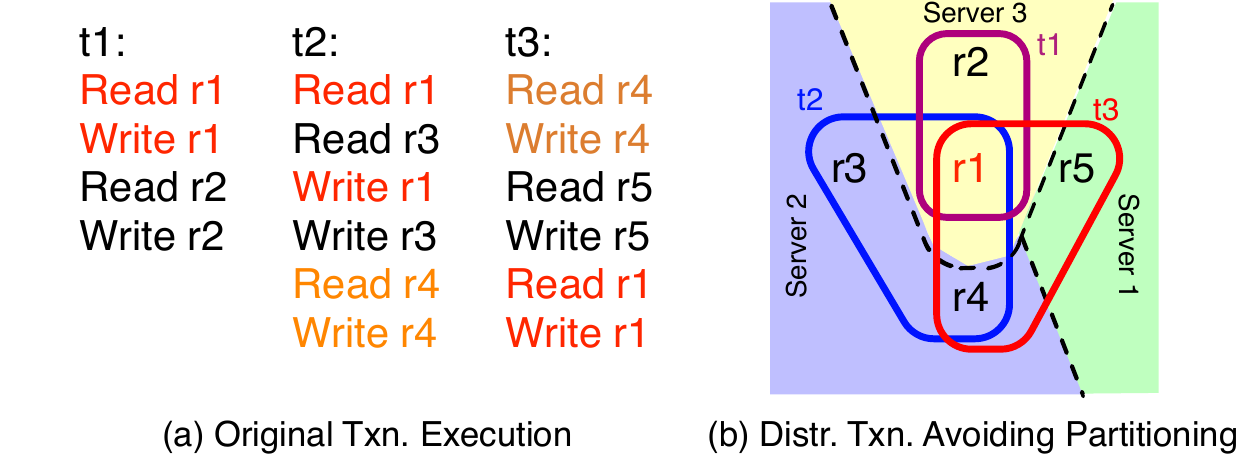}
        \caption{Traditional Execution and Partitioning.}
        \label{fig:trad_example}
    \end{minipage}
    \hspace{30pt}
    \begin{minipage}{.45\textwidth}
        \centering
        \includegraphics[width=\textwidth]{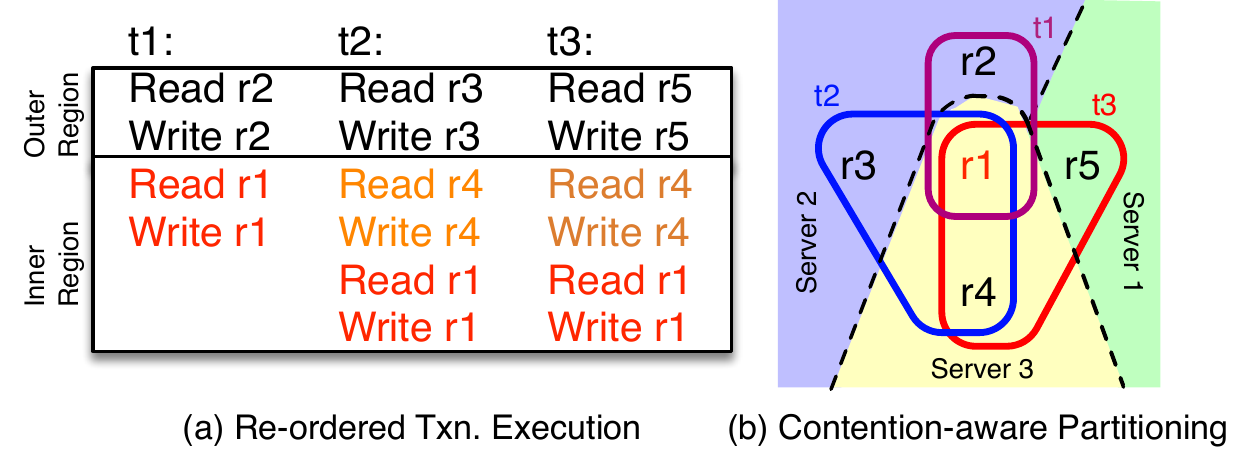}
        \caption{\system{} Execution and Partitioning.}
        \label{fig:chiller_example}
    \end{minipage}%
    \vspace{1em}
\end{figure*}

In this paper, we present \system{}, a new partitioning scheme and execution model based on 2-phase-locking which aims to minimize contention.
\system{} is based on two complementary ideas: {\bf (1) a novel commit protocol} based on re-ordering transaction operations with the goal of minimizing the lock duration for contended records through committing such records early, and {\bf (2) contention-aware partitioning} so that the most critical records can be updated without additional coordination.
For example, assume a simple scenario with three servers in which each server can store up to two records, and a workload consisting of three transactions $t_1$, $t_2$, and $t_3$ (Figure~\ref{fig:trad_example}a).
All transactions update $r_1$.
In addition, $t_1$ updates $r_2$, $t_2$ updates $r_3$ and $r_4$, and $t_3$ updates $r_4$ and $r_5$.
The common wisdom would dictate partitioning the data in a way that the number of cross-cutting transactions is minimized; in our example, this would mean co-locating all data for $t_1$ on a single server as shown in Figure~\ref{fig:trad_example}b, and having distributed transactions for $t_2$ and $t_3$.  

%[height=3.5cm]

However, as shown in Figure~\ref{fig:chiller_example}a, if we re-order each transaction's operations such that the updates to the most contended items ($r_1$ and $r_4$) are done last, we argue that it is better to place $r_1$ and $r_4$ on the same machine, as in Figure~\ref{fig:chiller_example}b.
At first this might seem counter-intuitive as it increases the total number of distributed transactions. 
However, this partitioning scheme decreases the likelihood of conflicts and therefore increase the total transaction throughput.
The idea is that re-ordering the transaction operations minimizes the lock duration for the ``hot'' items and subsequently the chance of conflicting with concurrent transactions.
More importantly, after the re-ordering, the success of a transaction relies entirely on the success of acquiring the lock for the most contended records. 
That is, if a distributed transaction has already acquired the necessary locks for all non-contended records (referred to as the {\em outer region}), the commit outcome depends solely on the contended records (referred to as the {\em inner region}).
This allows us to make all updates to the records in the {\em inner region} without any further coordination.
Note that this partitioning technique primarily targets high-bandwidth low-latency networks, which mitigates the two most common bottlenecks for distributed transactions: message overhead and limited network bandwidth. 

To provide such a contention-aware scheme, Chiller is based on two complementary ideas that go hand-in-hand: a contention-aware data partitioning algorithm and an operation-reordering execution scheme. 
First, different from existing partitioning algorithms that aim to minimize the number of distributed transactions (such as Schism~\cite{curino2010schism}), Chiller's partitioning algorithm explicitly takes record contention into account to co-locate hot records. 
Second, at runtime, \system{} uses a novel execution scheme which goes beyond existing work on re-ordering operations (e.g., QURO~\cite{yan2016leveraging}).
By taking advantage of the co-location of hot records, Chiller's execution scheme reorders operations such that it can release locks on hot records early and thus reduce the overall contention span on those records.
As we will show, these two complementary ideas together provide significant performance benefits over existing state-of-the-art approaches on various workloads.

In summary, we make the following contributions: 
\begin{enumerate}[label=(\textbf{\arabic*}),topsep=1pt,itemsep=0pt,parsep=0pt,leftmargin=15pt]
\item We propose a new contention-centric partitioning scheme.
\item We present a new distributed transaction execution technique, which aims to update highly-contended records without additional coordination.
\item We show that \system{} outperforms existing techniques by up to a factor of 2 on various workloads. 
\end{enumerate}
\section{Overview}
\label{sec:background}

The throughput of distributed transactions is limited by three factors: (1) message overhead, (2) network bandwidth, and (3) increased contention~\cite{binnig2016nam}.
The first two limitations are significantly alleviated with the new generation of high-speed RDMA-enabled networks.
However, what remains is the increased contention likelihood, as message delays are still significantly longer than local memory accesses.

\subsection{Transaction Processing with 2PL \& 2PC}
\label{sec:contention:traditional}

To understand the impact of contention in distributed transactions, let us consider a traditional 2PL with 2PC.
Here, we use transaction $t_3$ from Figure~\ref{fig:trad_example}, and further assume that its coordinator is on Server 1, as shown in Figure~\ref{fig:2pc-a}. 
The green circle on each partition's timeline shows when it releases its locks and commits.
We refer to the time span between acquisition and release of a record lock as the record's \textit{contention span} (depicted by thick blue lines), during which all concurrent accesses to the record would be conflicting.
In this example, the contention span for all records is 2 messages long with piggybacking optimization (when merging the last step of execution with the prepare phase) and 4 without it.

While our example used 2PL, other concurrency control (CC) methods suffer from this issue to various extents~\cite{harding2017evaluation}.
For example in OCC, transactions must pass a validation phase before committing.
If another transaction has modified the data accessed by a validating transaction, it has to abort and all its work will be wasted~\cite{dashti2017transaction, harding2017evaluation}.

\subsection{Contention-Aware Transactions}
\label{sec:contention:transactions}

\begin{figure}[t]
\centering
\begin{subfigure}[b]{0.50\columnwidth}
   \includegraphics[width=\textwidth]{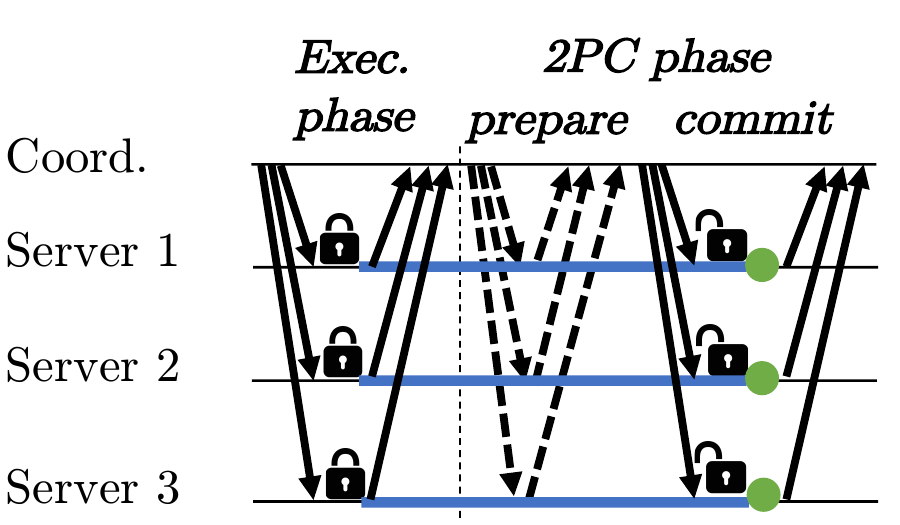}
   \caption{2PC + 2PL }
   \label{fig:2pc-a}
\end{subfigure}\hfill
\begin{subfigure}[b]{0.50\columnwidth}
   \includegraphics[width=\textwidth]{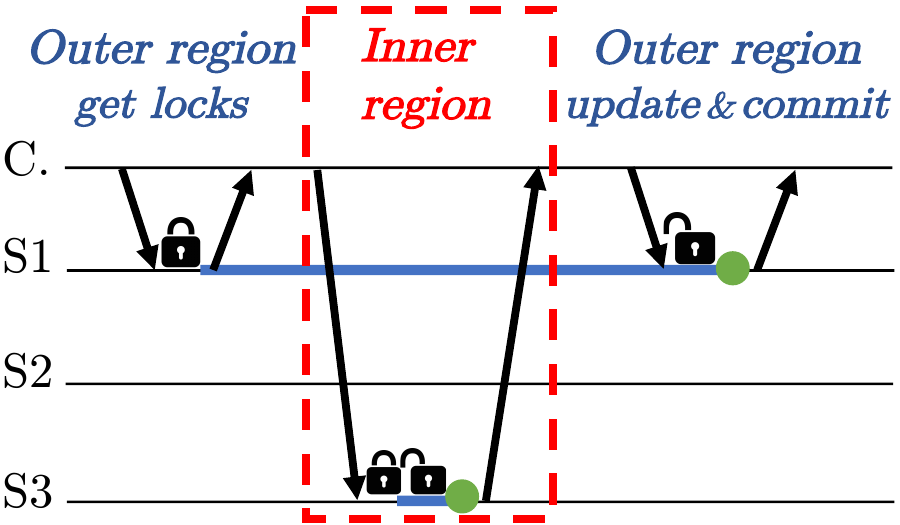}
   \caption{Two-region execution}
   \label{fig:2pc-b}
\end{subfigure}
\caption{The lifetime of a distributed transaction. The green dots denote when each server releases its locks. The blue lines represent the contention span for each server.}
\label{fig:2pc}
\end{figure}

We propose a new partition and execution scheme that aims to minimize the contention span for contended records.
The partitioning layout shown in Figure~\ref{fig:chiller_example}b opens new possibilities.
As shown in Figure~\ref{fig:2pc-b}, the coordinator requests locks for all the non-contended records in $t_3$, which is $r_5$.
If successful, it will send a request to the partition hosting the hot records, Server 3, to perform the remaining part of the transaction.
Server 3 will attempt to acquire the lock for its two records, complete the read-set, and perform the transaction logic to check if the transaction can commit.
If so, it \textbf{commits} the changes to its records.
% Server 3's commit point, therefore, happens earlier than the other two involved partitions. 

The reason that Server 3 can unilaterally commit or abort before the other involved partitions receive the commit decision is that Server 3 contains all necessary data to perform the transaction logic.
Therefore, the part of the transaction which deals with the hottest records is treated as if it were an independent \textit{local} transaction.
This effectively makes the contention span of $r_1$ and $r_4$ much shorter (just local memory access, as opposed to at least one network roundtrip).

\subsection{Discussion}
\label{sec:assumptions}
There are multiple details and simplifications hidden in the execution scheme presented above.

First, after sending the request to Server 3, neither the coordinator nor the rest of the partitions is allowed to abort the transaction and this decision is only up to Server 3.
For this reason, our system currently does not support triggers, which may cause the transaction to abort at any arbitrary point.
In that matter, its requirement is very similar to that of H-Store~\cite{kallman2008hstore}, VoltDB~\cite{stonebraker2013voltdb}, Calvin~\cite{thomson2012calvin} and MongoDB~\cite{banker2011mongodb}.
Also, the required determinism to disallow transactions to abort after a certain point in their life cycles is realized through the combination of Chiller's novel execution, replication and recovery protocols, which will be discussed in Section~\ref{sec:ft}.

Second, for a given transaction, the number of partitions for the inner region has to be \textit{at most} one.
Otherwise, multiple partitions cannot commit independently without coordination.
This is why executing transactions in this manner requires a new partitioning scheme to ensure that contended records that are likely to be accessed together are co-located.

Finally, our execution model needs to have access to the transaction logic in its entirety to be able to re-order its operations.
Our prototype achieves this by running transactions through invoking stored procedures, though it can be realized by other means such as implementing it as a query compiler (similar to Quro~\cite{yan2016leveraging}).
Due to the low overhead of our re-ordering algorithm, ad-hoc transactions can also be supported, as long as all operations of a transaction are issued in one shot.
The main alternative model, namely interactive transactions, in which there may be multiple back-and-forth rounds of network communication between a client application and the database is extremely unsuitable for applications that deal with contended data yet demand high throughput, because the database cannot reason about the boundaries of transactions upfront, and therefore all locks and latches have to be held for the entire scope of the client interaction which may last multiple roundtrips~\cite{thomson2013deterministic}.
\section{Two-region Execution}
\label{sec:transaction}

\begin{figure*}[h]
    \centering
    \includegraphics[width=1.0\textwidth]{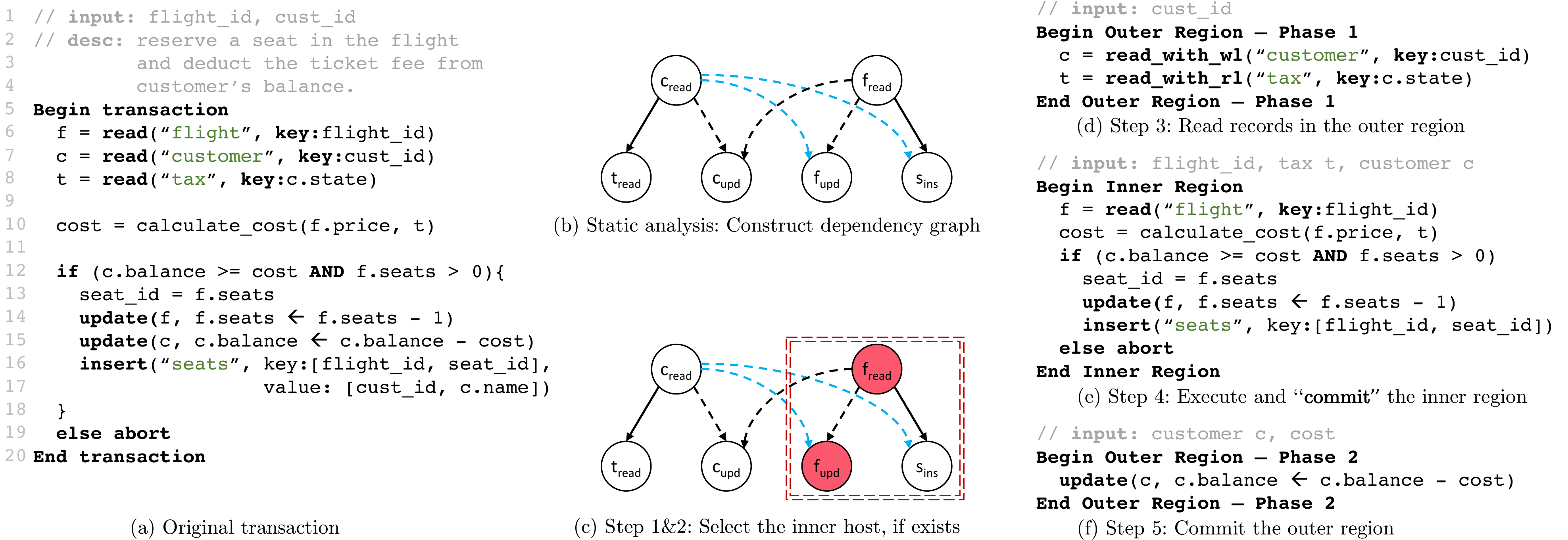}
    \caption{Two-region execution of a ticket purchasing transaction. In the dependency graph, primary key and value dependencies are shown in solid and dashed lines, respectively (blue for conditional constraints, e.g., an ``if'' statement).
    Assuming the flight record is contended (red circles), the red box in (c) shows the operations in the inner region (Step 4). The rest of the operations will be performed in the outer region (Steps 3 and 5).}
    \label{fig:transaction-processing}
\end{figure*}

\subsection{General Overview}
The goal of the two-region execution scheme is to minimize the duration of locks on contended records by postponing their lock acquisition until right before the end of the expanding phase of 2PL, and performing their lock release right after they are read/modified, without involving them in the 2PC protocol.
More specifically, the execution engine re-orders operations into cold operations (outer region) and hot operations (inner region); 
the outer region is executed as normal.
If successful, the records in the inner region are accessed.
The important point is that the inner region \textit{commits} upon completion without coordinating with the other participants.
Because of the way that the inner region is not involved in 2PC, fault tolerance requires a complete re-visit, otherwise many failure scenarios may sacrifice the system's correctness or liveness.
Until we discuss our fault tolerance algorithm in Section~\ref{sec:ft}, we present the execution and partitioning schemes under a no-failure assumption.

To help explain the concepts, we will use an imaginary flight-booking transaction shown in Figure~\ref{fig:transaction-processing}a.
Here, there are four tables: flight, customer, tax and seats.
In this example, if the customer has enough balance and the flight has an available seat (line 12), a seat is booked (lines 14 and 16) and the ticket fee plus state-tax is deducted from their account (line 15).
Otherwise, the transaction aborts (line 19).

The remainder of this section is structured as follows.
Section~\ref{sec:transaction:static} describes how we extract the constraints in re-ordering operations from the transaction logic and model it as a dependency graph.
Using this graph, a five-step protocol, described in Section~\ref{sec:transaction:runtime}, is used to execute a two-stage transaction.
Finally, in Section ~\ref{sec:transaction:discussion}, we present optimizations, and look at challenges for the protocol to be correct and fault tolerant.
Our solution to these challenges is then presented throughout the subsequent sections. 

\subsection{Constructing a Dependency Graph}
\label{sec:transaction:static}
There may be constraints on data values that must hold true (e.g., availability of a flight seat).
Furthermore, operations in a transaction may have dependencies among each other.
The goal is to reflect such constraints in the \emph{dependency graph}.
Here, we distinguish between two types of dependencies.
A \textit{primary key dependency (pk-dep)} is when accessing a record $r_2$ can happen only after accessing record $r_1$, as the primary key of $r_2$ is only known after $r_1$ is read (e.g., the read operation for the tax record in line 8 must happen after the read operation for the customer record in line 7).
In a \textit{value dependency (v-dep)}, the new values for the update columns of a record $r_2$ are known only after accessing $r_1$ (e.g. the update operation in line 15).
We are only concerned about the pk-deps, and not the v-deps, since v-deps do not restrict the order of lock acquisition, while pk-deps do put restrictions on which re-orderings are possible.

Each operation of the transaction corresponds to a node in the dependency graph.
There is an edge from node $n_1$ to $n_2$ if the corresponding operation of $n_2$ depends on that of $n_1$.
The dependency graph for our running example is shown in Figure~\ref{fig:transaction-processing}b.
For example, the insert operation in line 16 ($s_{\scriptsize\mbox{ins}}$ in the graph) has a pk-dep on the read operation in line 6 ($f_{\scriptsize\mbox{read}}$), and has a v-dep on the read operation in line 7 ($c_{\scriptsize\mbox{read}}$).
This means that getting the lock for the insert query can only happen after the flight record is read (pk-dep), but it can happen before the customer is read (v-dep).
Please refer to the figure's caption for the explanation of the color codes.

\subsection{Run-Time Decision}
\label{sec:transaction:runtime}
Given the dependency graph, we describe step-by-step how the protocol executes a two-region transaction.
 
\textbf{1) Decide on the execution model:}
First, it must find the list of candidate operations for the inner region.
An operation can be a candidate if the record(s) accessed by it is marked as contended in the lookup table, and it does not have any pk-dep on operations on other partitions, since if it does, it would make early commit of the inner region impossible. 
In Figure~\ref{fig:transaction-processing}b, if the insert operation $s_{\scriptsize\mbox{ins}}$ belongs to a different partition than $f_{\scriptsize\mbox{read}}$, the latter cannot be considered for the inner region because there is a pk-dep between them.

Finding the hosting partition of an operation which accesses records by their primary keys is quite straightforward. However, finding this information for operations which access records by non-primary-key attributes may require secondary indexes.
In case no such information is available, such operations will not be considered for the inner region.

\textbf{2) Select the host for inner region:}
If all candidate operations for the inner region belong to the same host, then it is chosen as the \emph{inner host}, otherwise, one has to be picked.
Currently, we choose the host with the highest number of candidate operations as the inner host.

\textbf{3) Read records in outer region:}
The transaction attempts to lock and read the records in its outer region. 
In our example, an exclusive lock for the customer record and a shared lock for the tax record are acquired.
If either of these lock requests fails, the transaction aborts.

\textbf{4) Execute and commit inner region:}
Once all locks have been acquired for the records in the outer region, the coordinator delegates processing the inner region to the inner host by sending a message with all information needed to execute its part (e.g., transaction ID, input parameters, etc.).
Having the values for all of the records in the read-set allows the inner host to check if all of the constraints in the transaction are met (e.g., there are free seats in the flight).
This guarantees that if operation in the outer region should result in an abort, it will be detected by the inner host and the entire transaction will abort.

Once all locks are successfully acquired and the transaction logic is checked to ensure the transaction can commit, the inner host updates the records, replicates its changes to its replicas (Section~\ref{sec:ft:replication}) and \textit{commits}.
In case any of the lock requests or transaction constraints fails, the inner host aborts the transaction and directly informs the coordinator about its decision.
In our example, the update to the flight record is applied, a new record gets inserted into the seats table, the partial transaction commits, and the value for the \texttt{cost} variable is returned, as it will be needed to update the customer's balance by another partition. 

\textbf{5) Commit outer region:}
If the inner region succeeds, the transaction is already considered committed and the coordinator must commit all changes in the outer region.
In our example, the customer's balance gets updated, and the locks are released from the tax and customer records.

\subsection{Challenges}
\label{sec:transaction:discussion}
First, it will not be useful if the hot records of a transaction are scattered across different partitions.
No matter which partition becomes the inner host, the other contended records will observe long contention spans.
Therefore, frequently co-accessed hot records must be co-located.
To this end, we present a novel partitioning technique in Section~\ref{sec:partitioning}.
Second, the inner host removes its locks earlier than the other participants (steps 4 and 5).
For this reason, fault tolerance requires a revisit, which will be presented in Section~\ref{sec:ft}.
\section{Contention-aware Partitioning}
\label{sec:partitioning}
\begin{figure*}[t]
\centering
\begin{subfigure}[b]{0.38\textwidth}
   \includegraphics[width=\textwidth]{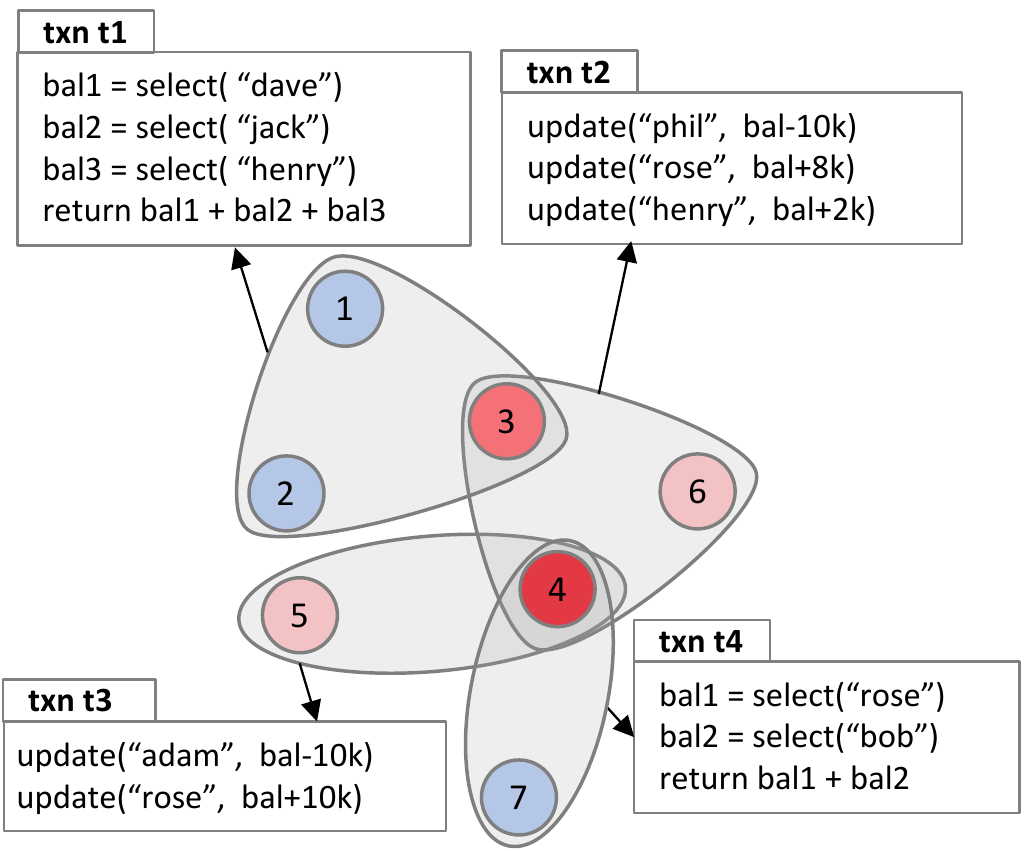}
   \caption{
   The original workload. The hyper edges show the transactions' boundaries. Darker red indicates more contended.
   }
   \label{fig:workload}
\end{subfigure}\hfill
\begin{subfigure}[b]{0.26\textwidth}
   \includegraphics[width=0.84\textwidth]{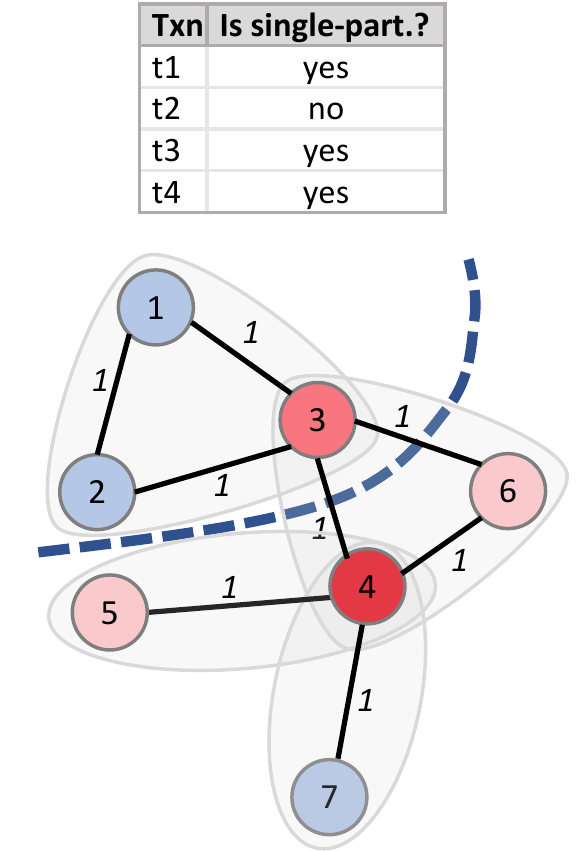}
   \caption{
   Distributed transaction minimization schemes. Edge weights are co-access frequencies.
   }
   \label{fig:schism}
\end{subfigure}\hfill
\begin{subfigure}[b]{0.26\textwidth}
   \includegraphics[width=0.84\textwidth]{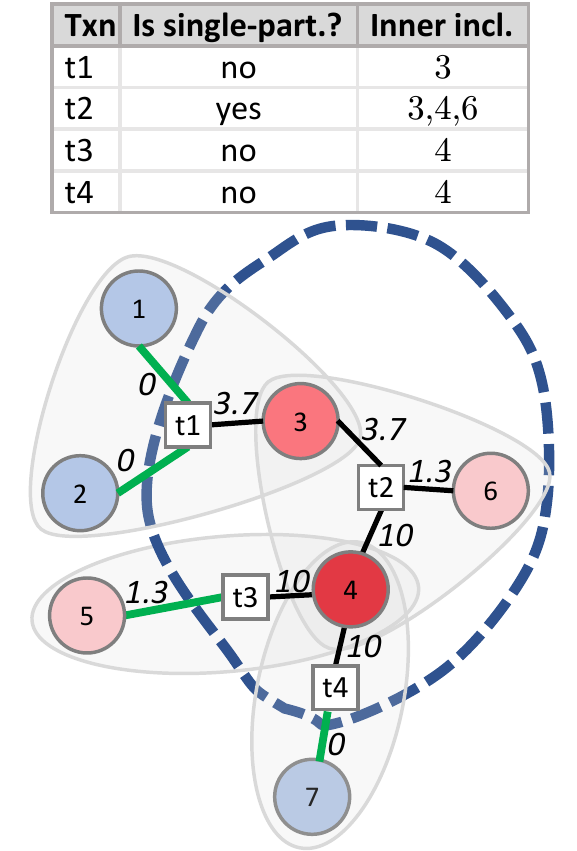}
   \caption{
   Contention-centric partitioning. Squares denote transactions. Green edges show outer regions.
   }
   \label{fig:chiller-contention-graph}
\end{subfigure}
\vspace{-2ex}
\caption{An example workload and how partitioning techniques with different objectives will partition it into two parts.}

\label{fig:example-workload}
   %\vspace{-3ex}
\end{figure*}

To fully unfold the potential of the two-region execution model, the objective of our proposed partitioning algorithm is to find a horizontal partitioning of the data which minimizes the contention.
To better explain the idea, we will use 4 transactions shown in Figure~\ref{fig:example-workload}.
The shade of red corresponds to the record hotness (darker is hotter), and the goal is to find two balanced partitions (for now, we define ``balanced'' as a partitioning that splits the set of records in half).
Existing partitioning schemes, such as Schism~\cite{curino2010schism} minimize distributed transactions, as shown in Figure~\ref{fig:schism}.
However, such a split would increase the contention span for records $3$ or $4$, and $6$ in transaction $t_2$, because $t_2$ will have to hold locks on either $3$ or $4$, and $6$ as part of an outer region.

Not only the main objective of our proposed partitioning is different than the one in existing techniques, but also their commonly used graph representation of the workload (with records as vertices and co-accesses as edges) does not capture the essential requirements of our problem, that is, the distinction of inner and outer region and differences in their execution.
This necessitates a new workload representation.

\subsection{Overview of Partitioning}
\label{sec:partitioning:overview}
To measure the hotness of records, servers randomly sample the transactions' read- and write-sets during execution.
These samples are aggregated over a pre-defined time interval by the \textit{partitioning manager} server (PM).
PM uses this information to estimate the contention of individual records (Section~\ref{sec:partitioning:contention}).
It then creates the graph representation of the workload which accommodates the requirements for the two-region execution model (Section~\ref{sec:partitioning:graph}).
Based on this representation, it then uses a graph partitioning tool to partition the records with the objective of minimizing the overall contention of the workload (Section~\ref{sec:partitioning:algorithm}).
Finally, it updates servers' lookup tables with new partition assignments.

We assume henceforth that the unit of locking and partitioning is records.
However, the same concepts apply to more coarse grained lock units, such as pages or hash buckets.

\subsection{Contention Likelihood}
\label{sec:partitioning:contention}
Using the aggregated samples, PM calculates the conflict likelihood for each record.
More specifically, we define the probability of a conflicting access for a given record as:
\vspace{-2ex}
\[
P_c(X_w, X_r) = \overbrace{P(X_w > 1)P(X_r = 0)}^{(i)} + \overbrace{P(X_w > 0)P(X_r > 0)}^{(ii)}
\]
\vspace{-3ex}

Here, $X_w$ and $X_r$ are random variables corresponding to the number of times a given record is read or modified within the lock window, respectively.
The equation consists of two terms to account for the two possible conflict scenarios: (i) write-write conflicts, and (ii) read-write conflicts.
Since (i) and (ii) are disjoint, we can simply add them together. 

Similar to previous work~\cite{yu1993ontheanalytical,kraska2009consistency}, we model $X_w$ ($X_r$) using a Poisson process with a mean arrival time of  $\lambda_w$  ($\lambda_r$), which is the time-normalized access frequency.
This allows us to rewrite the above equation as follows: 
\vspace{-2ex}
\[
\begin{split}
    P_c(X_w, X_r) &=
    \overbrace{
    \Big(1 - (\frac{ {\lambda_w}^0 e^{-\lambda_w} }{0!} + \frac{ {\lambda_w}^1 e^{-\lambda_w} } {1!}) \Big)
    \Big( \frac{ {\lambda_r}^0 e^{-\lambda_r} }{0!} \Big)
    }^{(i)} \\
    & \qquad +  
    \overbrace{
    \Big( 1-\frac{ {\lambda_w}^0 e^{-\lambda_w} }{0!} \Big) 
    \Big( 1-\frac{ {\lambda_r}^0 e^{-\lambda_r} }{0!} \Big)
    }^{(ii)}
    % \\ &=
    % \Big(1-e^{-\lambda_w}(1 + \lambda_w) \Big)
    % \Big(e^{-\lambda_r} \Big) \\
    % & \qquad +
    % \Big( 1-e^{-\lambda_w} \Big)
    % \Big( 1-e^{-\lambda_r} \Big)
    \\ &=
    1 - e^{-\lambda_w} - \lambda_w e^{-\lambda_w} e^{-\lambda_r}
\end{split}
\]
\vspace{-3ex}

The contention likelihood is defined per lock unit.
We use $P_c(\rho)$ to refer to the contention likelihood of record $\rho$.
In the equation above, when $\lambda_w$ is zero, meaning no write has been made to the record, $P_c(\rho)$ will be zero, since shared locks are compatible so no conflict is expected.
With a non-zero $\lambda_w$, higher values of $\lambda_r$ will increase the contention likelihood due to the conflict of read and write locks.

\subsection{Graph Representation}
\label{sec:partitioning:graph}
The are three key properties that a graph representation of the workload should have to properly fit in the context of our execution model.
First, record contentions must be captured in the graph as this is the main objective.
Second, the relationship between records must also be modeled, due to the requirement that there can be only one inner region for a transaction, and hence the frequently co-accessed contended records should be co-located.
Third, the final partitioning should also make it possible to determine the inner region for each transaction.
Therefore, \system{} models the workload quite differently than existing partitioning algorithms.

As shown in Figure~\ref{fig:chiller-contention-graph}, we model each transaction as a star; at the center is a dummy vertex (referred to as \textit{t-vertex}, denoted by squares) with edges to all the records that are accessed by that transaction.
Thus, the number of vertices in the graph is $|T| + |R|$, where $|T|$ is the number of transactions and $|R|$ is the number of records.
The number of edges will be the sum of the number of records involved per transaction.

All edges connecting a given record-vertex (\textit{r-vertex}) to all of its t-vertex neighbors have the same weight.
This weight is proportional to the record's contention likelihood.
The weight of the edge between an r-vertex and a connected t-vertex reflects how bad it would be if the record is not accessed in the inner region of that transaction.

Note that while our graph representation does not directly incorporate dependencies among operations (e.g., \textit{pk-dep}), it should not take long for a running system until the partitioning algorithm would automatically build edges between records frequently accessed as part of these operations.

Applying the contention likelihood formula to our running example and normalizing the weights will produce the graph with the edge weights in Figure~\ref{fig:chiller-contention-graph}.
Note that there is no edge between any two records. 
Co-accessing records is implied by a common t-vertex connecting them.
Next, we describe how our partitioning algorithm takes this graph as input and generates a partitioning with low contention.

\subsection{Partitioning Algorithm}
\label{sec:partitioning:algorithm}

As we are able to model contention among records using a weighted graph, we can apply standard graph partitioning algorithms.
More formally, our goal is to find a partitioning, which minimizes the contention: 

\vspace{-4ex}
\begin{gather*}
    \min_{S} \sum_{\rho \in R} P_{c}^{(S)}(\rho) \\
    s.t. \;\; \forall p \in S : L(p) \leqslant (1 + \epsilon) \cdot \mu
\end{gather*} 
\vspace{-4ex}

Here, $S$ is a partitioning of the set of records $R$ into $k$ partitions,  $P_{c}^{(S)}(\rho)$ is the contention likelihood of record $\rho$ under partitioning $S$, $L(p)$ is the load on partition $p$, $\mu$ is the average load on each partition, and $\epsilon$ is a small constant that controls the degree of imbalance.
Therefore, $\mu = \frac{\sum_{p \in P} L(p)}{|P|}$.
The definition of load will be discussed shortly.

Chiller makes use of METIS~\cite{Karypis1998metis}, a graph partitioning tool which aims to find a high-quality partitioning of the input graph with a small cut, while at the same time respecting the constraint of approximately balanced load across partitions.

The interpretation of the partitioning is as follows:
A cut edge $e$ connecting a r-vertex $v$ in one partition to a t-vertex $t$ in another partition implies that $t$ will access $v$ in its outer region, and thus observing a conflicting access with a probability proportional to $e$'s weight.
To put it differently, the partition to which $t$ is assigned determines the inner host of $t$; all r-vertices assigned to the same partition can be executed in the inner region of $t$.
As a result, a split that minimize the total weight of all cut edges also minimizes the contention. 

In our example, the sum of the weights of all cut edges (which are portrayed as green lines) is $1.3$.
Transaction $t_1$ will access record $3$ in its inner region as its t-vertex is in the same partition as record $3$, while it will access records $1$ and $2$ in its outer region.
Even though the number of multi-partition transactions is increased compared to the partitioning in Figure~\ref{fig:schism}, this split results in a much lower contention.

The load $L$ for a partition can be defined in different ways, such as the number of executed transactions, hosting records, or record accesses.
The vertex weights depends on the chosen load metric.
For the metric of number of executed transactions, t-vertices have a weight of 1 while r-vertices will have a weight of 0.
The weighting is reversed for the second metric.
METIS generates a partitioning such that the sum of vertex weights in each partition is approximately balanced.

\subsection{Discussion}
\label{sec:partitioning:optimizations}

\subsubsection{Scalability of Partitioning.}
\label{sec:partitioning:scalability}
There are two issues every partitioning scheme has to address: (1) the graph size and the cost of partitioning it, and (2) the size of the lookup table.

It is time- and computation-intensive to partition very large graphs.
However, \system{} has a unique advantage over existing partitioning techniques: it produces graphs with significantly fewer edges for most workloads. 
Schism, for instance, introduces a total of $n(n-1) / 2$ edges for a transaction with $n$ records~\cite{curino2010schism}. 
However, Chiller's star representation introduces only $n$ edges per transaction, resulting in a much smaller graph.
For example, we found that on average, constructing the workload graph and applying the METIS partitioning tool take up to 5 times longer on Schism compared to Chiller in our experiments.

Furthermore, our approach provides a unique opportunity to reduce the size of the lookup table. 
As we are mainly interested in reducing contention, we can primarily focus on the records with a contention likelihood above a given threshold. 
Hence, the lookup table only needs to store where these hot records are located.
The other records can be partitioned using an orthogonal scheme (e.g., hash or range), which takes no lookup-table space. 
We study this technique in more depth using an experiment (Section~\ref{sec:evaluation:lookup}).

\subsubsection{Re-Partitioning and Data Migration.}
\label{sec:partitioning:repart}
While the process described in Section~\ref{sec:partitioning:overview} can be done periodically for the purpose of re-partitioning, our current prototype is based on an offline implementation of the Chiller partitioner.
In our experiments, running this algorithm on 100 thousand sampled transactions for a workload with as many as 30 millions records took less than ten minutes on one machine (Section~\ref{sec:evaluation:setup}).
This time includes calculating the contention likelihoods, building the workload graph, and partitioning it using METIS.
In addition, we found that the pruning techniques proposed above are quite effective in reducing the partitioning time without significantly impacting the throughput. 
Therefore, we envision that the offline re-partitioning scheme would be sufficient for many workloads, and re-partitioning can be as simple as running the algorithm on one or multiple partitioning managers.
For other workloads with more frequently changing hot spots, however, it is possible that constantly relocating records in an incremental way is more effective.
Extending this work to support online re-partitioning is an interesting direction of future work.

Another related topic is when data re-partitioning happens, how the system relocates records while still maintaining ACID guarantees.
Our current prototype produces a record relocation list, which can be used to move records transactionally (each tuple migration is performed as one individual transaction).
As data migration is a general requirement by every production OLTP partitioning tool, there are many automatic tools which perform this task more efficiently~\cite{elmore2011zephyr,wei2017replication,elmore2015squall}.
We are planning to extend our prototype to use Squall~\cite{elmore2015squall}, which is a live data migration tool.

\subsubsection{Minimizing Distributed Transactions. }
It is also possible to co-optimize for contention and distributed transactions using the same workload representation.
One only needs to assign a minimum positive weight to all edges in the graph.
The bigger the minimum weight, the stronger the objective to co-locate records from the same transaction.
Such co-optimization is still relevant even in fast RDMA-enabled networks since a remote access through RDMA is about $10\times$ slower than a local access.
However, as we argue in this paper, the optimal partitioning objective should shift in the direction of minimizing contention, and therefore minimizing distributed transactions is just a secondary optimization.
\section{Fault Tolerance}
\label{sec:ft}
The two-region execution model presented in Section~\ref{sec:transaction} modifies the typical 2PC for transactions accessing contended records.
A transaction is considered committed once its processing is finished by the inner host, after which, it \textit{must} be able to commit on the other participants even despite failures.
A participant in the outer region cannot unilaterally abort a transaction once it has granted its locks in the outer region and informed the coordinator, since the transaction may have already been committed by the inner host.

Chiller employs write-ahead logging to non-volatile memory.
However, similar to 2PC, while logging enables crash recovery, it does not provide high availability.
The failure of the inner host before sending its decision to the coordinator may sacrifice the availability, since the coordinator would not know if the inner region is already committed or not, in which case it has to wait for the inner host to recover.

To achieve high availability, Chiller relies on a new replication method based on synchronous log-shipping, explained in Section~\ref{sec:ft:replication}.
Then, we discuss how this replication protocol achieves high availability while still maintaining consistency.

\subsection{Replication Protocol}
\label{sec:ft:replication}
In conventional synchronous log-shipping replication, the logs are replicated before the transaction commits.
Since in Chiller, the transaction commit point (i.e., when the inner region commits) happens \textit{before} the outer region participants commit their changes (see Figure~\ref{fig:2pc-b}), the inner region replication cannot be postponed to the end of the transaction, otherwise its changes may be lost if the inner host fails.

\begin{figure}[t]
    \centering
    \includegraphics[width=1\columnwidth]{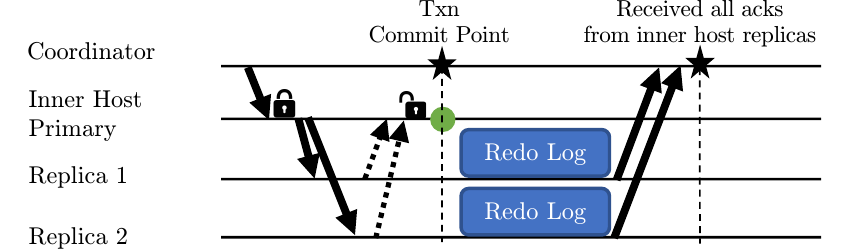}
    \caption{Replication algorithm for the inner region.}
    \label{fig:inner-replication}
\end{figure}

To solve this problem, Chiller employs two different algorithms for the replication of the inner and outer regions.
The changes in the outer region are replicated as normal---once the coordinator finishes performing the operations in the transaction, it replicates the changes to the replicas of the outer region before making the changes visible.
The inner region replication, however, must be done \textit{before the transaction commit point}, so that the commit decision will survive failures.
Below, we describe the inner region replication in terms of the different roles of the participants:
\\
\textbf{Inner host}:
As shown in Figure~\ref{fig:inner-replication}, when the inner host finishes executing its part, it sends an RPC message to its replicas containing the new values of its records, the transaction read-set, and the sequence ID of the replication message.
It then waits for the acks from its NIC hardware, which guarantee that the messages have been successfully sent to the replicas.
Finally, it safely commits its changes.
In case the inner host decides to abort, the replication phase will not be needed and it can directly reply to the coordinator.
\\
\textbf{Inner host replicas}:
Each replica applies the updates in the message in the sequence ID order.
This guarantees that the data in the replicas synchronously reflect the changes in the primary inner host partition.
When all updates of the replication message are applied, each replica notifies the \textit{original coordinator} of the transaction, as opposed to responding back to the inner host.
This saves one network message delay.
\\
\textbf{Coordinator}: 
The coordinator is allowed to resume the transaction only after it has received the notifications from all the replicas of the inner host.
 
In the following, we describe our failure recovery protocol.

\subsection{Failure Recovery}
\label{sec:ft:recovery}
For failure detection and membership reconfiguration, Chiller relies on a cluster manager such as Zookeeper~\cite{hunt2010zookeeper}.
When a machine is suspected of failure, all of the other nodes close their communication channels to the suspected node to prevent it from interfering in the middle of the recovery process.

The recovery procedure is as follows:
First, each partition $p$ probes its local log, and compiles a list of pending transactions on $p$.
For each transaction, its coordinator, inner host, the list of outer region participants are retrieved, and then aggregated at a designated node to make a global list of pending transactions.
Below, possible failure scenarios for a pending two-region transaction along with how the fault tolerance is achieved are discussed.
\\
\textbf{Failure of inner host}:
If none of the surviving replicas of a failed inner host has received the replication message, the transaction can be safely aborted, because it indicates that the inner host has not committed either.
However, if at least one of its replicas has received such a message, that transaction \textit{can} commit, even though that it might have not yet replicated on all the replicas.
In this case, the coordinator finishes the transaction on the remaining inner host replicas and the outer region participants, and commits.
\\
\textbf{Failure of coordinator}:
If a node is found to be the inner host (or one of its replicas, in case the inner host is failed too), it will be elected as the new coordinator, since it already has the values for the transaction read-set.
Otherwise, the transaction can be safely aborted because its changes are not yet received/committed by its inner host.
\\
\textbf{Failure of an outer region participant}:
If the failure of participant $i$ happens before the coordinator initiates the inner region, then the transaction is safely aborted.
Otherwise, one of $i$'s replicas which has been elected as the new primary will be used to take over $i$'s role in the transaction.

\vspace{0.2em}
\textbf{Proof of Correctness} --- We now provide some intuition on the protocol correctness in terms of safety and liveness.

It is easy to see why the two-region execution model with the described replication protocol maintains safety, since transactions are serialized at the point when their inner host commits.
Also, similar to 2PC, if even one participant commits (aborts), no other participant is allowed to abort (commit).
This is due to the ``no turning back'' concept of the commit protocol of the inner region, guaranteeing that all participants will agree on the same decision.

To support liveness, the system first needs to detect failures and repair/replace faulty nodes.
For this purpose, Chiller relies on the existence of a fault tolerant coordinator, such as Zookeeper~\cite{hunt2010zookeeper} or Chubby~\cite{burrows2006chubby}.
So long as at most $f$ out of $f+1$ replicas fail, the protocol guarantees liveness by satisfying these two properties: \\
(1) A transaction will eventually commit at all its participants and their replicas once it has been committed by the inner host:
The inner host commits only when its changes are replicated on its replicas.
Therefore, there will be at least one replica containing the commit decision which causes the transaction to commit during the recovery process. \\
(2) A transaction which is not yet committed at its inner host will eventually either abort or commit:
If the inner host does not fail, it will eventually process the inner region.
However, if it encounters a failure, the transaction will be discovered during the failure recovery, and handled by a new inner host. \\
\section{Implementation}
\label{sec:system}
We now briefly present the implementation of the system we used to evaluate Chiller.

In our system, tables are built on top of a distributed hash table, and are split horizontally into multiple partitions, with each \textit{execution server} hosting one partition in its main memory.
The unit of locking is a hash bucket, and each hash bucket encapsulates its lock metadata in its header, eliminating the need for a centralized lock manager per partition.
Our current implementation performs locking on bucket granularity and does not prevent the phantom problem.
However, the core idea of Chiller also works with range locks.

An execution server utilizes all its processing cores through multi-threading, where each execution thread has access to the hosted partition on that machine.
To minimize inter-thread communication, each transaction is handled from beginning to end by one execution thread.
To extract maximum concurrency, each worker thread employs multiple \textit{co-routine workers}, such that when one transaction is waiting for a network operation, it yields to the next co-routine worker which processes a new transaction.
The communication needed for distributed transactions is done either through direct remote memory operations (RDMA Read, Write, and atomic operations), or via RPC messages implemented using RDMA Send and Receive.
The access type for each table is specified by the user when creating that table.

Bucket to partition mappings are stored in a lookup table, which is replicated on all servers so that execution threads would know where each record is.
It can be either defined by the user in the form of hash or range functions on some table attributes, or produced individually for all or some buckets using the partitioning algorithm.
In addition to storing the partition assignments, the lookup table also contains the list of buckets with a contention above a given threshold, which is used to determine the inner region for each transaction.

As discussed in Section~\ref{sec:ft}, to guarantee high availability in the presence of failures, we use log-shipping with 2-safety, where each record is replicated on two other servers.
The backup replicas get updated synchronously before the transaction commits on the primary replica.
In addition, like other recent high-performance OLTP systems~\cite{dragojevic2015farm,kalia2016fasst,zamanian2017namdb}, crash recovery is guaranteed by relying on the persistence of execution servers' logs on some form of non-volatile memory (NVM), such as battery-backed DRAM.
\begin{figure*}[t]
\centering
\includegraphics[width=0.55\textwidth]{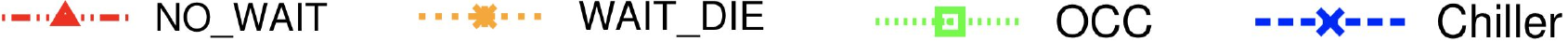}\\
\begin{subfigure}[b]{0.31\textwidth}
    \captionsetup{margin={0.5cm, 0cm}}
    \includegraphics[width=\textwidth]{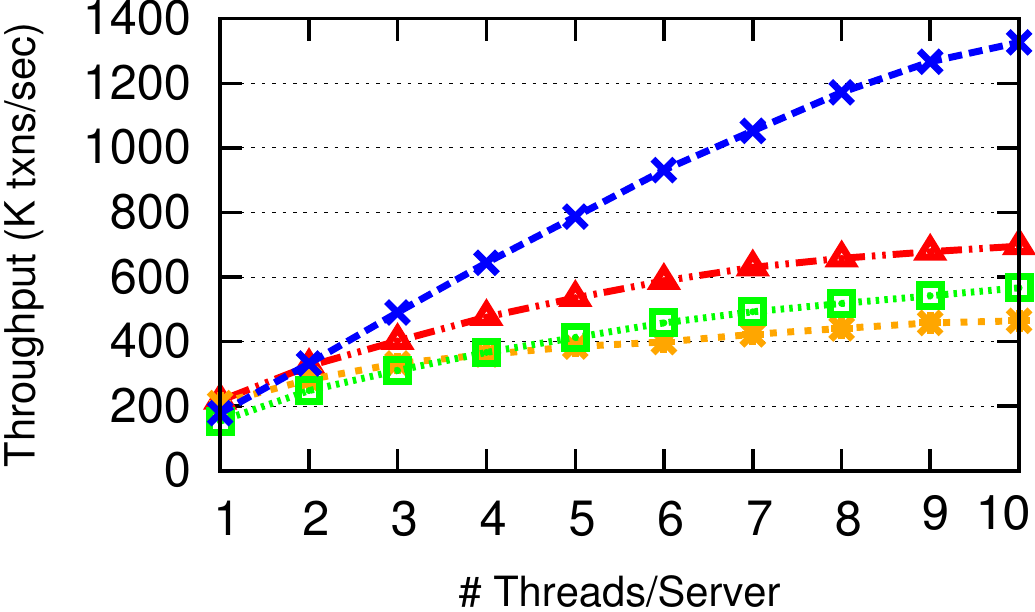}
    \caption{Throughput for standard TPC-C.}
    \label{fig:exp-threads:throughput}
\end{subfigure}\hspace{0.3cm}
\begin{subfigure}[b]{0.31\textwidth}
    \captionsetup{margin={0.5cm, 0cm}}
   \includegraphics[width=\textwidth]{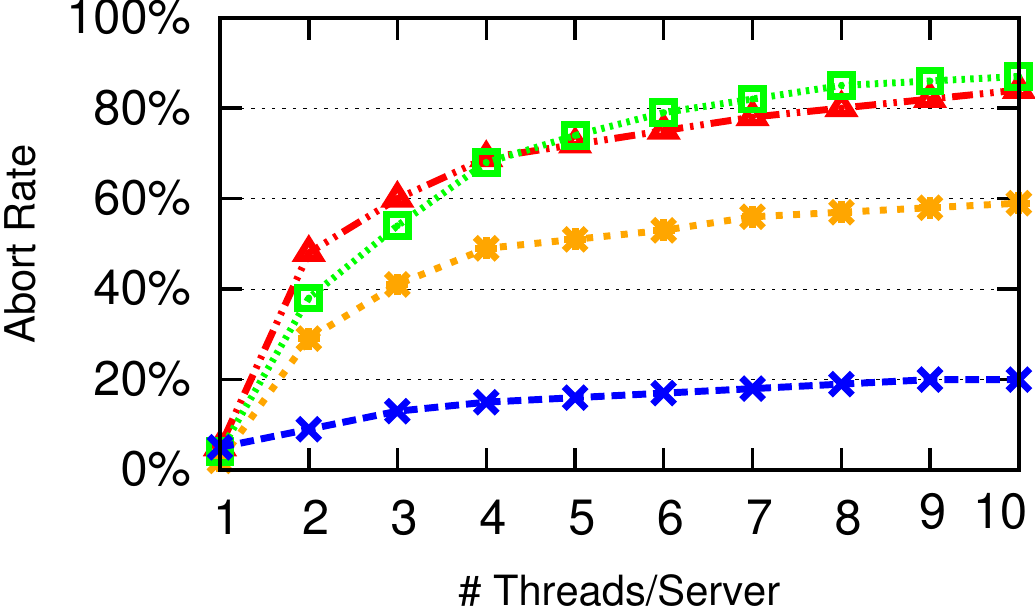}
   \caption{Abort rate for standard TPC-C.}
   \label{fig:exp-threads:abort} 
\end{subfigure}\hspace{0.3cm}
\begin{subfigure}[b]{0.31\textwidth}
    \captionsetup{margin={0.5cm, 0cm}}
    \includegraphics[width=\textwidth]{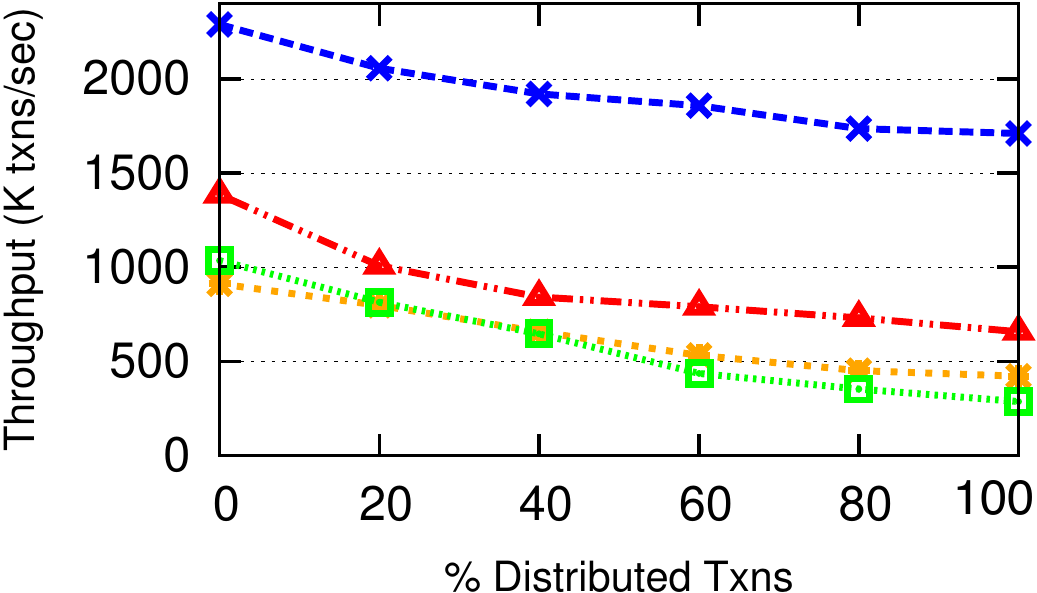}
    \caption{Impact of distributed transactions.}
    \label{fig:exp-distributed-txn}
\end{subfigure}
\caption{Comparison of different concurrency control methods and Chiller for the standard and modified TPC-C.}
\label{fig:exp-threads}
\end{figure*}

\section{Evaluation}
\label{sec:evaluation}

We evaluated our system to answer two main questions:

\begin{enumerate}[label=(\textbf{\arabic*}),topsep=1pt,itemsep=0pt,parsep=0pt,leftmargin=15pt]
\item How does \system{} and its two-region execution model perform under various levels of contention compared to existing techniques?
\item Is the contention-aware data partitioning effective in producing results that can efficiently benefit from the two-region execution model?
\end{enumerate}

\subsection{Setup}
\label{sec:evaluation:setup}

The test bed we used for our experiments consists of 7 machines connected to a single InfiniBand EDR 4X switch using a Mellanox ConnectX-4 card.
Each machine has 256GB RAM and two Intel Xeon E5-2660 v2 processors with 2 sockets and 10 cores per socket.
In all experiments, we use only one socket per machine where the the NIC is directly attached and disable hyper-threading to minimize the variability in measurements caused by same-core threads interference, which is a typical setup also used in other papers \cite{sirin2017methodology,yuan2016bcc,zamanian2019rethinking}.
The machines run Ubuntu 14.04 Server Edition as their OS and Mellanox OFED 3.4-1 driver for the network.

\subsection{Baselines}
To assess the ability of the two-region execution model in handling contention, we evaluate how it holds up against alternative commonly used concurrency control (CC) models, more specifically these protocols:

\textbf{Two-Phase Locking (2PL):} we implemented two widely used variants of distributed 2PL with deadlock prevention.
In \textbf{\nowait{}}, the system aborts a transaction once it suspects of a deadlock, i.e., when a record lock request is denied. 
Therefore, waiting for locks is not allowed.
In \textbf{\waitdie{}}, transactions are assigned unique timestamps before execution.
An older transaction is allowed to wait for a lock which is owned by a younger transaction, otherwise it aborts.
Timestamp ordering ensures no deadlock is possible. 
While one could also implement 2PL with deadlock detection, it demands significant network synchronization between servers to detect lock-request cycles, and is therefore very costly in a distributed setting~\cite{harding2017evaluation, appuswamy2017analyzing}.
Therefore, we did not include it in our evaluation. 

We based the implementation of Chiller's locking mechanism on \nowait{} due to its lower overhead (no need to manage lock queues), although \waitdie{} could also be used.

\textbf{Optimistic (OCC):} 
We based our implementation on the MaaT protocol~\cite{mahmoud2014maat}, which is an efficient and scalable algorithm for \occ{} in distributed settings~\cite{harding2017evaluation}.
Each transaction is assigned a range for its commit timestamp, initially set to $\big[0$ $\infty \big)$.
Also, the DBMS stores for each record the list of pending reader IDs and writer IDs, and the ID of the last committed transaction which accessed the record.  
Each time a transaction reads/modifies a record, it modifies its timestamp range to be in compliance with the read/write timestamp of that record, and adds its unique timestamp to the list of the record's read/write IDs.
At the end, each participant of the transaction attempts to validate its part by changing the timestamp ranges of the validating transaction and the other conflicting transactions, and votes to commit if the final range of the transaction is valid.
The coordinator commits a transaction only if all the participants vote to commit.

In addition, we evaluate two common partitioning schemes:\\
\textbf{Hash-partitioning} is the method of assigning records to partitions based on the hash value of their primary key(s).\\
\textbf{Schism} is the most notable automatic partitioning technique for OLTP workloads.
It first uses Metis to find a small cut of the workload graph, then compares this record-level partitioning to both a decision tree-learned range partitioning and a simple hash partitioning and picks the one which results in the minimum number of distributed transactions, or if equal, requires a smaller lookup table.
We include the results for different CC schemes for Schism partitioning, and report only \nowait{} for hash partitioning as a simple baseline.

\subsection{Workloads}
\label{sec:evaluation:workload}

For our experiments, we use the following workloads and measure throughput as number of committed transactions per second (i.e., excluding aborted transactions).

\textbf{TPC-C:}
This is the de facto standard for evaluating OLTP systems.
It consists of 9 tables and 5 types of transactions.
The majority of transactions access records belonging to a single warehouse.
Therefore, the obvious partitioning layout is by warehouse.
Despite being highly partitionable, it contains two severe contention points. 
First, each \textit{new-order} transaction does an increment on one out of $10$ records in the district table of a warehouse.
Second, every \textit{payment} transaction updates the total balance of a warehouse and one of its $10$ districts, creating an even more severe contention point.
These two transactions comprise more than $87\%$ of the workload.
We used one warehouse per server (i.e., 7 warehouses in total) which translates to a high contention workload.
This allows us to focus on the differences in the execution models of Chiller and traditional schemes.

\textbf{YCSB:}
It consists of a single table with 1KB records~\cite{cooper2010benchmarking}.
We generated 5 million records ($\sim5$ GB) per server.
To generate read and write-sets of transactions with a desired level of locality, we used a mapping function from records to partitions.
Since the benchmark does not specify transactions, we group multiple read/write-operations into one transaction as discussed next.
To explore different aspects of the problem in more depth, we used the following two workloads:\\
\textit{YCSB Local:} This workload represents a perfectly partitionable dataset.
Each transaction reads and modifies 16 records stored on a single partition using a Zipfian distribution with varying skew factor $\theta$.
\\
\textit{YCSB Distributed:} Many real OLTP workloads are not as partitionable as \textit{YCSB Local} on the transaction level, but still exhibit some locality on the record level.
For example, a purchase that contains one Harry Potter book is likely to contain a few other volumes of the Harry Potter franchise, while still including any other non-related item.
To model such cases, we generated a workload where each transaction reads 4 records across different partitions of the entire database uniformly, and reads and modifies 2 other records from a single partition using a Zipfian distribution.

\textbf{InstaCart: }
To assess the effectiveness of our approach to deal with difficult to partition workloads, we used a real-world data set released by Instacart~\cite{instacart2017}, which is an online grocery delivery service.
The dataset contains over $3$ million grocery orders for around $50$K items from more than $200$K of their customers.
On average, each order contains $10$ grocery products purchased in one transaction by a customer.
To model a transactional workload based on the Instacart data, we used the TPC-C's \texttt{NewOrder} where each transaction reads the stock values of a number of items, subtracts each one by 1, and inserts a new record in the order table.
However, instead of randomly selecting items according to the TPC-C specification, we used the actual Instacart data set. 
Unlike the original TPC-C, this data set is actually difficult to partition due to the nature of grocery shopping, where items from different categories (e.g., dairy, produce, and meat) may be purchased together.
More importantly, there is a significant skew in the number of purchases of different products.
For example, $15\%$ of transactions contain banana.

\subsection{TPC-C Results}
As common in all TPC-C evaluations, all tables are partitioned by warehouse ID, except for the Items table which is read-only and therefore replicated on all servers.
Both Chiller and Schism produce this partitioning given the workload trace, therefore in the following experiments, we mainly focus on the two-region execution feature of Chiller, and evaluate it against the other CC schemes. 

\subsubsection{Impact of Concurrency Level.}
We first measure the performance of Chiller, \nowait{}, \waitdie{}, and \occ{} with increasing number of worker threads per server.
Although such increase provides more CPU power to process transactions, it also increases the contention.
Studying this factor is therefore of great importance since many modern in-memory databases are designed for systems with multi-core CPUs.

As Figure~\ref{fig:exp-threads:throughput} shows, with only one worker thread running in each machine (i.e., no concurrent data access), \nowait{} and \waitdie{} perform similarly, and has $10\%$ higher throughput than Chiller.
This is accounted by the two-region execution overhead.
However, as we increase the number of worker threads, the likelihood that transactions conflict with each other increases, negatively impacting the scalability of 2PL and \occ{}.
\system{}, on the other hand, minimizes the lock duration for the two contention points in TPC-C (warehouse and district records) and thus, scales much better.
With 10 threads, the throughput of Chiller is $2\times$ and $3\times$ higher than that of \nowait{} and \waitdie{}, respectively.

Figure~\ref{fig:exp-threads:abort} shows the corresponding abort rates (averaged over all threads).
With more than 4 threads, \occ{}'s abort rate is even higher than \nowait{}, which is attributed to the fact that many transactions are executed to the validation phase and then are forced to abort.
Compared to the other techniques, the abort rate of \system{} increases much more slowly as the level of concurrency per server increases.

This experiment shows the inherent scalability issue with traditional CC schemes when deployed on multi-core systems, and how Chiller manages to significantly alleviate it.  

\subsubsection{Impact of Distributed Transactions.}

For this experiment, we restricted the transactions to \texttt{NewOrder} and \texttt{Payment}, each making up $50\%$ of the mix (In the standard TPC-C workload, these two transactions are the only ones which can be multi-partition).
For \texttt{Payment}, we varied the probability that the paying customer is located at a remote warehouse, and for \texttt{NewOrder} we varied the probability that at least one of the purchased items is located in a remote partition.

Figure~\ref{fig:exp-distributed-txn} shows the total throughput with a varying fraction of distributed transactions.
As the percentage of distributed transactions increases, the already existing conflicts become more pronounced due to the prolonged duration of transactions, since a higher ratio of transactions must wait for network roundtrips to access records on remote partitions.
This observation clearly shows why having good partitioning layout is a necessity for good performance in traditional CC protocols, and why existing partitioning techniques aim to minimize the percentage of distributed transactions.

Also, compared to the traditional concurrency protocols, \system{} degrades the least when the fraction of distributed transactions increases.
More specifically, the performance of \system{} drops only by $26\%$, while \nowait{} and \waitdie{} both observe close to $50\%$ drop in throughput, and the throughput of \occ{} has the largest decrease, which is about $73\%$. 
This is because the execution threads for a partition always have useful work to do; when a transaction is waiting for remote data, the next transaction can be processed.
Since in \system{}, conflicts are handled sequentially in the inner region, concurrent transactions have a much smaller likelihood of conflicting with each other.
Therefore, an increase in the percentage of distributed transactions only means higher latency per transaction, and not much increased contention, therefore has much less impact on the throughput.
This highlights our claim that minimizing the number of multi-partition transactions should not be the primary goal in the next generation of OLTP systems that leverage fast networks, but rather that optimizing for contention should be.

\subsection{YCSB Results}
\begin{figure*}[t]
\centering
\includegraphics[width=0.68\textwidth]{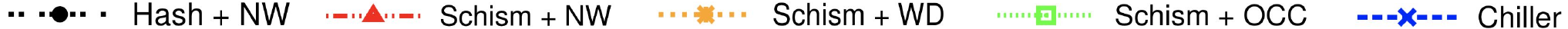}
\begin{subfigure}[b]{0.42\textwidth}
    \captionsetup{margin={0.5cm, 0cm}}
    \includegraphics[width=\textwidth]{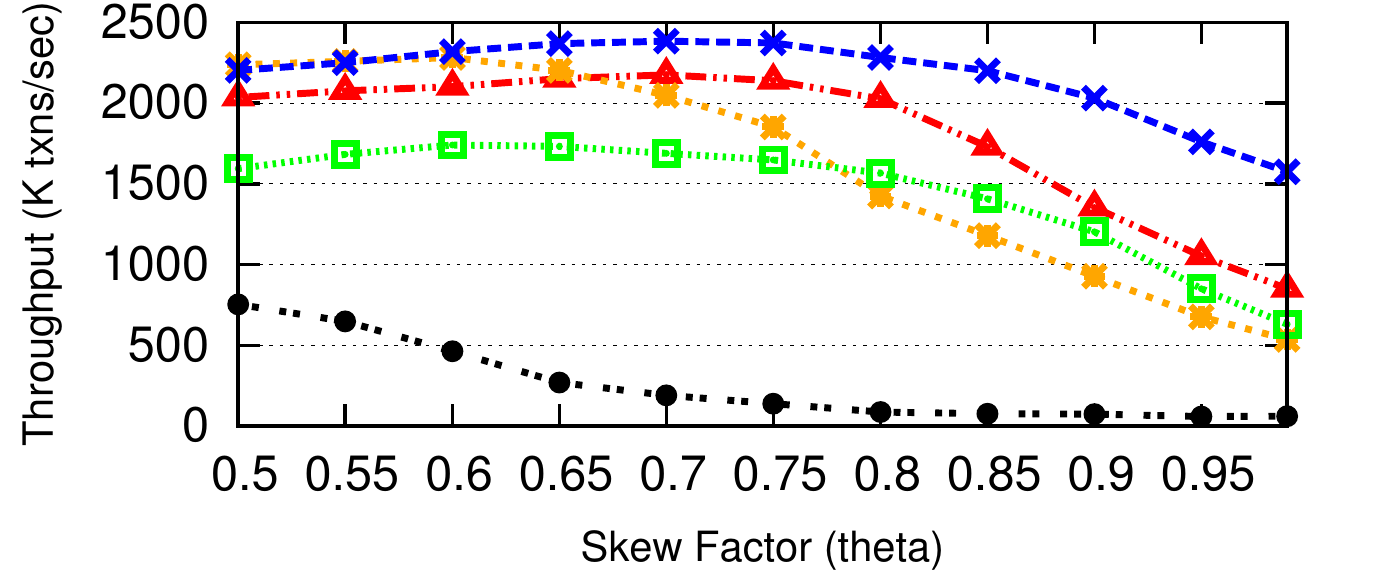}
    \caption{Total Throughput.}
    \label{fig:exp-ycsb-local:thruput}
\end{subfigure}\hfill
\begin{subfigure}[b]{0.42\textwidth}
    \captionsetup{margin={0.5cm, 0cm}}
   \includegraphics[width=\textwidth]{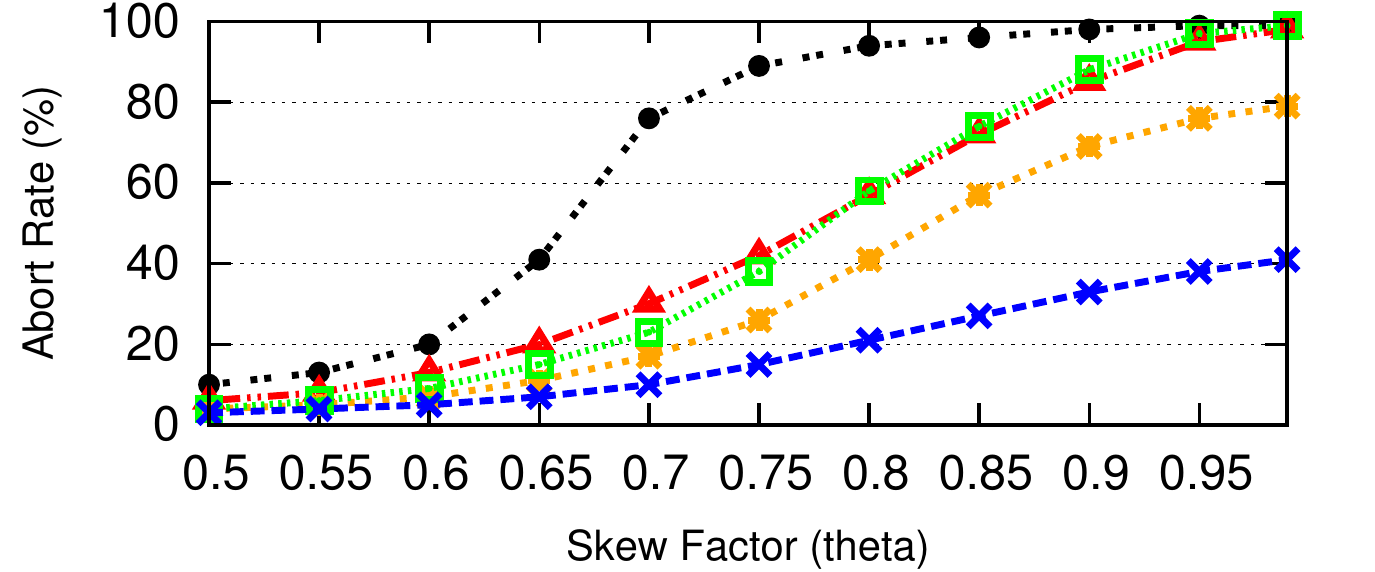}
   \caption{Average Abort Rate.}
   \label{fig:exp-ycsb-local:abort}
\end{subfigure}
\caption{\textbf{YCSB local} (all single-partition transactions).}
\label{fig:exp-ycsb-local}
\end{figure*}

\subsubsection{Single-Partition Transactions.}

We begin by examining the impact of contention on single-partition transactions.
We use the \textit{YCSB local} workload and vary the skew level $\theta$ from 0.5 (low skew) to 0.99 (high skew, the default in the original YCSB).
The aggregated throughput and the average abort rate are shown in Figures~\ref{fig:exp-ycsb-local:thruput} and \ref{fig:exp-ycsb-local:abort}.
For this workload, both Chiller and Schism can produce the same split as the ground truth mapping function we used to generate transactions.
As explained before, under traditional CC schemes, distributed transactions significantly intensify any contention in the workload, which explains the steep increase in the abort rate of the hash partitioning baseline in Figure~\ref{fig:exp-ycsb-local:abort}.

As the contention increases, all traditional CC schemes face high abort rate, reaching more than $50\%$ with $\theta=0.85$.
Chiller, on the other hand, is able to minimize the contention and hence reduce the abort rate.
When the skew is high ($\theta=0.99$), the throughput of Chiller is more than $85\%$ higher than the second best baseline, \nowait{}, while its abort rate is about half of \waitdie{}.
The initial increase in the throughput of Chiller and Schism-based schemes can be attributed to cache effects: with higher contention, there is a smaller working set of data for most transactions, making CPU/NIC caching more effective. However, as the contention further increases, the higher abort rate outweighs the caching effect.

This experiment shows that even for a workload with only single-partition transactions which is considered the sweet spot of the traditional partitioning and CC techniques, high contention can result in a major performance degradation, and Chiller's two-region execution model manages to alleviate the problem to a great extent.

\subsubsection{Scalability of Distributed Transactions.}
\label{sec:evaluation:distr}
\begin{figure}[t]
\centering
    \includegraphics[width=0.80\columnwidth]{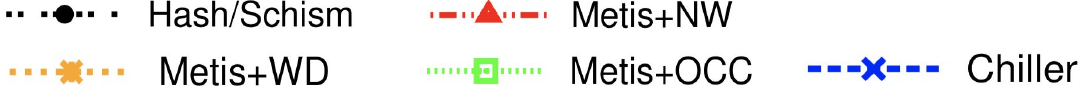}
    \includegraphics[width=0.41\textwidth]{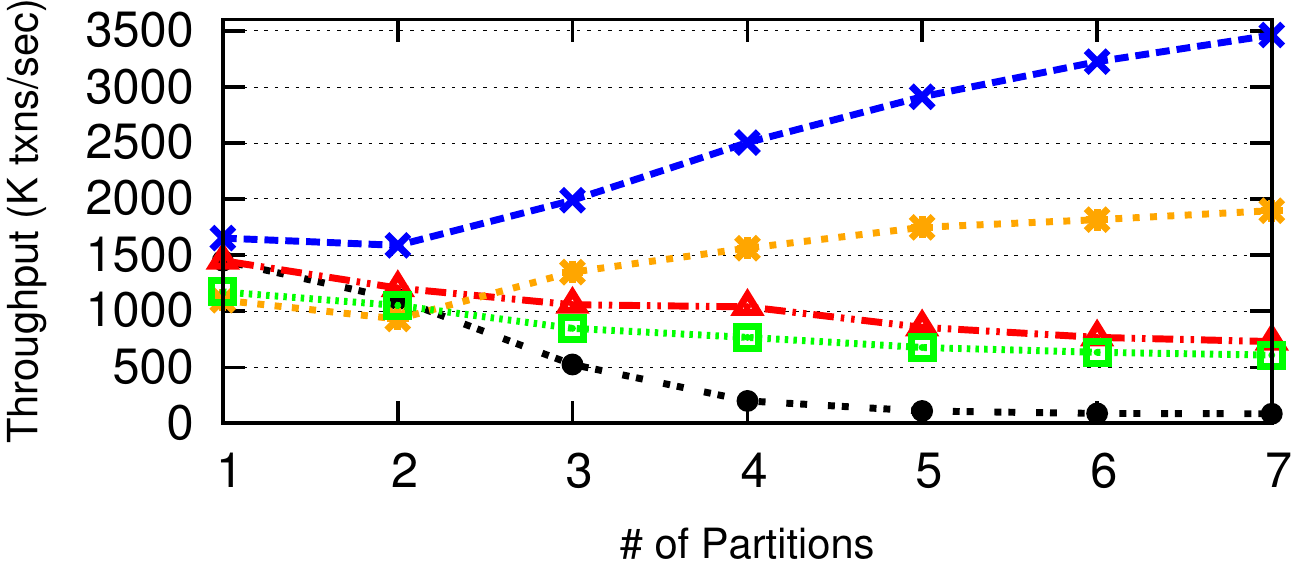}
	\caption{\textbf{YCSB distributed} with increasing cluster sizes.}
    \label{fig:exp-ycsb-distr}
\end{figure}

We next compare the scalability of different schemes in the presence of distributed transactions.
For this purpose, we used \textit{YCSB distributed} workload, in which each transaction reads 4 records from the entire database, and modifies 2 skewed records from the same partition.
Schism gives up fine-grained record-level partitioning and chooses simple hash partitioning, because in this workload, co-locating the hot records is not advantageous to simple hashing in terms of minimizing distributed transactions.
Therefore, we also show the results for the original partitioning produced by Metis, which aims to minimize the number of cross-partition record accesses.

Figure ~\ref{fig:exp-ycsb-distr} shows the throughput of the different protocols as the number of partitions (and servers) increases.
To avoid cluttering the graph, we only show the performance of the best CC scheme for Schism, which is \nowait{}.
To maintain a consistent replication factor of two, for the cluster size of one and two, we dedicate two and one extra backup machines, respectively, which do not process new transactions, and only process replication logs.

At first, all schemes drop in throughput despite the increase in the number of machines from one to two, which is due to the introduced distributed transactions.
Surprisingly, as the number of partitions increases, all the CC schemes which use Metis partitioning outperforms the one which uses Schism, even though that almost all transactions are distributed in both cases.
This is because in Metis, the contended records are co-located, and this drastically reduces the negative impact of aborting transactions in our system.
More specifically, a transaction which fails to get one of the two contended locks would release the other one immediately, whereas in the partitioning produced by Schism, these two records are likely to be placed in different partitions, and releasing the locks for an aborted transaction may take one network roundtrip, further intensifying the contention problem.
\waitdie{} performs better than \nowait{} since its waiting mechanism is able to overcome the lock thrashing issue in \nowait{}, though we note that we also observed this phenomenon for \waitdie{} for workloads where transactions access a higher number of hot records (e.g. see Figure~\ref{fig:exp-ycsb-local:abort}).
Compared to all the other schemes, Chiller scales much better since not only its partitioning co-locates the contended records together, but also its two-region execution model is able to access those records in inner regions of transactions and therefore significantly reduces the contention.
In fact, our measurements showed that the abort rate of Chiller with 7 machines is only $11\%$, whereas it is $81\%$ and $86\%$ for \nowait{} and \occ{}, respectively.
\waitdie{} again resulted in much lower abort rate of $49\%$, resulting in better scalability compared to the other baselines.
On 7 machines, the performance of Chiller is $4\times$ and $40\times$ of \nowait{} on Metis and Schism partitions, respectively, while close to $2\times$ of the second best baseline, \waitdie{}.

\subsubsection{Lookup Table Size.}
\label{sec:evaluation:lookup}

\begin{figure}[t]
\centering
    \includegraphics[width=0.85\columnwidth]{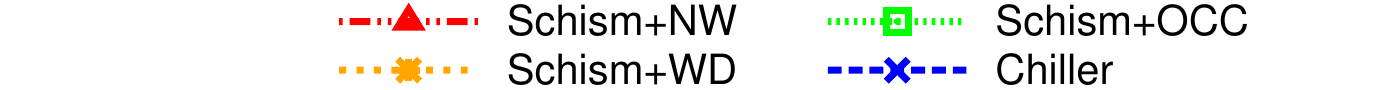}
    \includegraphics[width=0.85\columnwidth]{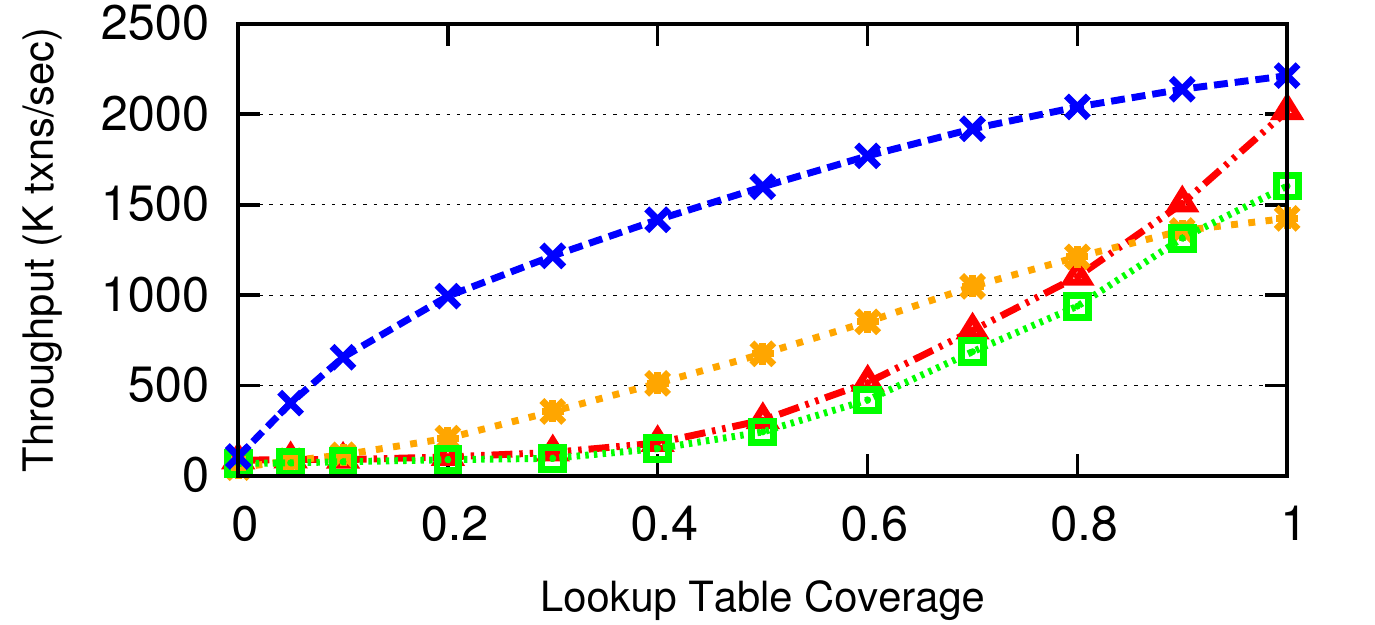}
	\caption{Varying the database coverage of the lookup table.}
    \label{fig:exp-lookup}
\end{figure}

In this experiment, we investigate the performance of our proposed scheme in situations where a full-coverage lookup table cannot be either obtained or stored, as discussed in Section~\ref{sec:partitioning:scalability}.
This can mainly happen when the number of database records is too large to store a complete lookup table on each machine.

We used \textit{YCSB local}, and fixed the skew parameter $\theta$ to $0.8$ to represent a moderately skewed workload.
Since all transactions in this workload are single-partition with respect to the ground truth mapping, both Schism and Chiller are able to find the optimal partitioning which makes all transactions single-partition, but this requires to cover the entire database in their resulting lookup tables.
To measure the impact of lookup table coverage, we vary the percentage of the records which are partitioned according to the optimization goal of each partitioning algorithm.
We used hash partitioning for the remaining records which, as a result, do not take up any lookup table entry, but result in a significant increase in the number of multi-partition transactions.

The results are shown in Figure~\ref{fig:exp-lookup}.
When all records are hash partitioned (the lookup table is empty), Chiller and all the other schemes achieve similar throughput.
As the lookup table increases in size, Chiller starts to diverge from the other schemes.
With a coverage of only $20\%$, Chiller achieves close to half of its peak throughput, whereas \nowait{} and \occ{} achieve less than $0.1$ of their peak throughput.
In contrast, \nowait{} relies on a $80\%$ coverage to achieve half of its max throughput. 
The wide gap between Chiller and the other protocols is due to the way that Chiller handles contention.
Placing the contended records (which are often a small fraction of the entire database) in the right partitions and handling them in inner regions are enough to remove most of the contention in the workload.
The rest of the records can be randomly assigned to partitions without increasing contention.

This experiment supports our claim in Section~\ref{sec:partitioning:optimizations} that, compared to partitioning schemes aiming to minimize distributed transactions, Chiller requires a much smaller lookup table to achieve a similar throughput.
In addition, while for this particular workload there exists a database split where each transaction accesses one partition, for many real workloads such partitioning does not exist.
Therefore, this experiment shows how Chiller compares against the other schemes for workloads with different degrees of partitionability.

\subsection{Instacart Results}
\label{sec:evaluation:instacart}
In our final experiment, we analyze the benefits of combining the Chiller's partitioning scheme with the two-region execution model.
We use a real-world Instacart workload (as introduced in Section \ref{sec:evaluation:workload}), which is harder to partition than TPC-C and YCSB.
Furthermore, we use the same replication factor of $2$ as for the previous experiments.

In order to understand whether or not the two-region execution model of Chiller is beneficial for the overall performance, we compare full Chiller (\text{Chiller}) to Chiller partitioning without the two-region execution model (\text{ChP}) and Chiller partitioning using Quro* (\text{ChP+Quro*}).
In contrast to \text{ChP} which does not re-order operations, \text{ChP+Quro*} re-orders operations using Quro~\cite{yan2016leveraging}, which is a recent contention-reduction technique for centralized database systems.
Moreover, we compare full Chiller to two other non-Chiller baselines (\text{Hash}-partitioning and \text{Schism}-partitioning).
For both \text{ChP} and \text{ChP+Quro*} as well as the non-Chiller baselines (\text{Hash} and \text{Schism}), we only show the results for a \waitdie{} scheme as it yielded the best throughput compared to \nowait{} and \occ{} for this experiment.

\begin{figure}[t]
\centering
    \includegraphics[width=0.65\columnwidth]{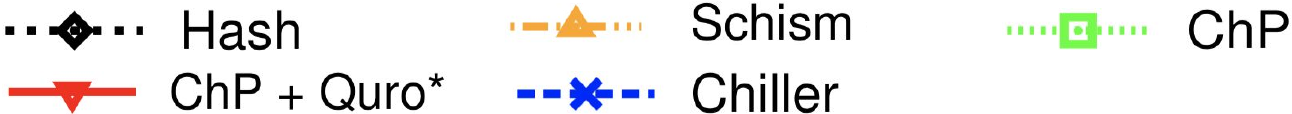}
    \includegraphics[width=0.4\textwidth]{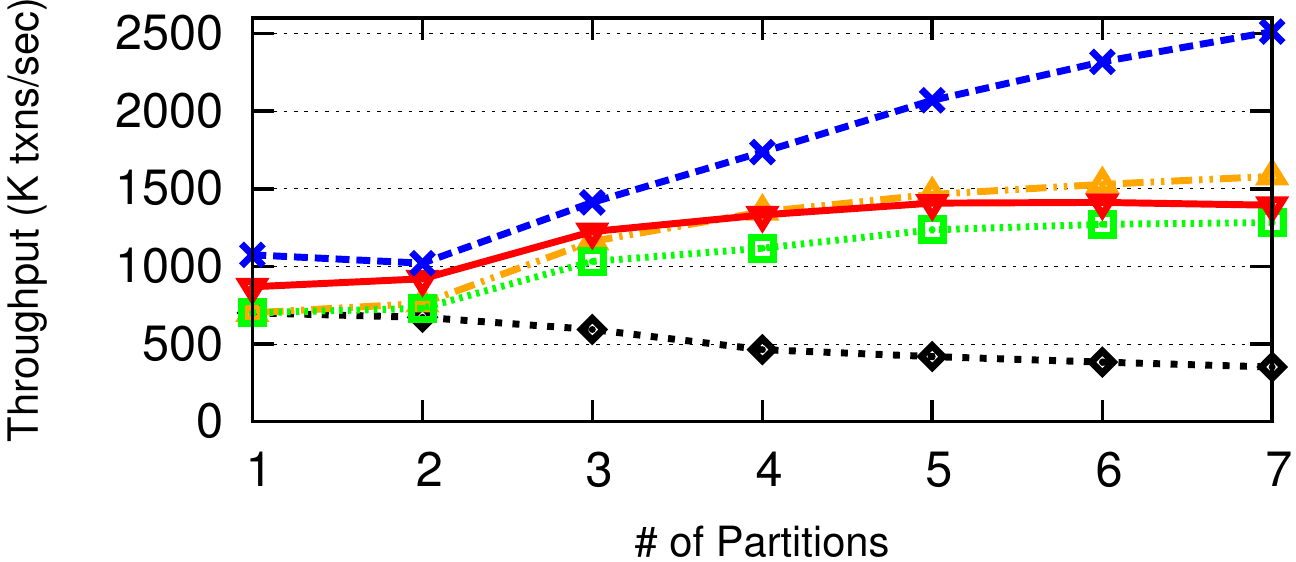}
	\caption{Instacart with different execution models.}
    \label{fig:exp-instacart}
\end{figure}

Figure~\ref{fig:exp-instacart} shows the results of this experiment for increasing cluster sizes.
Compared to the \text{Hash}-partitioning baseline (black line), both \text{ChP} and \text{ChP+Quro*} (green and red lines) have significantly higher throughput.
We found that this is not because the Chiller partitioning technique reduces the number of distributed transactions, but rather because contended records which are accessed together are co-located, which in turn reduces the cost of aborting transactions.
More specifically, if a transaction on contented records needs to be aborted, it only takes one round-trip, leading to an overall higher throughput since the failed transaction can be restarted faster (cf. Section~\ref{sec:evaluation:distr}).

Furthermore, we see that \text{ChP+Quro*}, which re-orders operations to access the low contended records first, initially increases the throughput by $20\%$ compared to \text{ChP} but then its advantage decreases as the number of partitions increases.
The reason for this is that the longer latency of multi-partition transactions offsets most of the benefits of operation re-ordering if the commit order of operations remains unchanged.
In fact, with 5 partitions, Schism (yellow line) starts to outperform \text{ChP+Quro*}, even though Schism does not leverage operation re-ordering.

In contrast to these baselines, Chiller (blue line) not only re-orders operations but also splits them into an inner and outer region with different commit points, and thus can outperform all the other techniques.
For example, for the largest cluster size, the throughput of Chiller is by approximately $1$ million txns/sec higher than the second best baseline. 
This clearly shows that the contention-centric partitioning must go hand-in-hand with the two-region execution to be most effective.
\balance{}
\section{Related Work}
\label{sec:related}

\textbf{Data Partitioning:}
A large body of work exists for partitioning OLTP workloads with the ultimate goal of minimizing cross-partition transactions~\cite{curino2010schism, tran2014jecb}. 
Most notably, Schism~\cite{curino2010schism} is an automatic partitioning and replication tool that uses a trace of the workload to model the relationship between the database records as a graph, and then applies METIS~\cite{Karypis1998metis} to find a small cut while approximately balancing the number of records among partitions.
Clay~\cite{serafini2016clay} builds the same workload graph as Schism, but instead takes an incremental approach to partitioning by building on the previously produced layout as opposed to recomputing it from scratch.
E-store~\cite{taft2014store} balances the load in the presence of skew in tree-structured schemas by spreading the hottest records across different partitions, and then moving large blocks of cold records to the partition where their co-accessed hot record is located.
Given the schema of a database, Horticulture~\cite{pavlo2012skew} heuristically navigates its search space of table schemas  to find the ideal set of attributes to partition the database. 
As stated earlier, all of these methods share their main objective of minimizing inter-partition transactions, which in the past have been known to be prohibitively expensive.
However, in the age of new networks and much ``cheaper'' distributed transactions, such an objective is no longer optimal.

\textbf{Transaction Decomposition:}
There has been also work exploring the opportunities in decomposing transactions into smaller units.
Gemini~\cite{li2012making} introduces a mixed consistency model called BlueRed in which transaction operations are divided into blue operations, which are eventually consistent with lower latency, and red operations, which are strongly consistent which require global serialization.
Gemini optimizes for overall latency and requires data to be replicated at all servers, and therefore does not have the notion of distributed transactions.
Chiller, on the other hand, optimizes for minimizing contention, and supports distributed transactions.
There has also been work on the theory of transaction chopping~\cite{shasha1995transaction,shasha1992simple,zhang2013transaction}, in which the DBMS splits a transaction into smaller pieces and treats them as a sequence of independent transactions.
In contrast to the idea of transaction chopping, our two-region execution not only splits a transaction into cold and hot operations, but re-orders operations based on which region they belong to.
Also, we do not treat the outer region as an independent transaction and will hold the locks on its records until the end of the transaction.
This allows us to our technique to abort a transaction later in the inner region.
Transaction chopping techniques, however, must adhere to \textit{rollback-safety}, in which all operations with the possibility of rollback must be executed in the first piece, since subsequent pieces must never fail. This restricts the possible ways to chop the transaction.

\textbf{Determinism and Contention-Reducing Execution:}
Another line of work aims to reduce contention through enforcing determinism to part or all of the concurrency control (CC) unit~\cite{cowling2012granola, kallman2008hstore, thomson2012calvin}.
In Granola~\cite{cowling2012granola}, servers exchange timestamps to serialize conflicting transactions.
Calvin~\cite{thomson2012calvin} takes a similar approach, except that it relies on a global agreement scheme to deterministically sequence the lock requests.
Faleiro et al.~\cite{faleiro2014lazy, faleiro2017high} propose two techniques for deterministic databases, namely lazy execution scheme and early write visibility, which aim to reduce data contention in those systems.
All of these techniques and protocols require \emph{a priori} knowledge of read-set and write-set. 

There has also been a large body of work on optimizing and extending traditional CC schemes to make them more apt for in-memory databases.
MOCC~\cite{wang16mostly} targets thousand-core systems with deep memory hierarchies and proposes a new concurrency control which mixes OCC with selective pessimistic read locks on contended records to reduce clobbered reads in highly contended workloads.
Recent work on optimistic CC leverages re-ordering operations inside a batch of transactions to reduce contention both at the storage layer and validation phase~\cite{ding18improving}.
While Chiller also takes advantage of operation re-ordering, it does so at an intra-transaction level without relying on transaction batching.
MV3C~\cite{dashti2017transaction} introduces the notion of repairing transactions in MVCC by re-executing a subset of a failed transaction logic instead of running it from scratch.
Most related to \system{} is Quro~\cite{yan2016leveraging}, which also re-orders operations inside transactions in a centralized DBMS with 2PL to reduce lock duration of contended data.
However, unlike \system{}, the granularity of contention for Quro is tables, and not records.
Furthermore, almost all these works deal with single-node DBMSs and do not have the notion of distributed transactions, 2PC or asynchronous replication on remote machines, and hence finding a good partitioning scheme is not within their scopes.

\textbf{Transactions over Fast Networks:}
This paper continues the growing focus on distributed transaction processing on new RDMA-enabled networks~\cite{binnig2016nam}.
The increasing adoption of these networks by key-value stores~\cite{mitchell2013using, kalia2015herd, li2017kv} and DBMSs~\cite{dragojevic2015farm, zamanian2017namdb, kalia2016fasst, chen2017fast}  is due to their much lower overhead for message processing using RDMA features, low latency, and high bandwidth.
These systems are positioned in different points of the spectrum of RDMA.
For example, FaSST~\cite{kalia2016fasst} uses the unreliable datagram connections to build an optimized RPC layer, and FaRM~\cite{dragojevic2015farm} and NAM-DB~\cite{zamanian2017namdb} leverage the RDMA feature to directly read or write data to a remote partition.
Though different in their design choices, scalability in the face of cross-partition transactions is a common promise of these systems, provided that the workload itself does not impose contention.
Therefore, \system{}'s two-region execution and its contention-centric partition are specifically suitable for this class of distributed data stores.
\section{Conclusions}
\label{sec:conclusion}
This paper presents \system{}, a distributed transaction processing and data partitioning scheme that aims to minimize contention.
\system{} is designed for fast RDMA-enabled networks where the cost of distributed transactions is already low, and the system's scalability depends on the absence of contention in the workload.
Chiller partitions the data such that the hot records which are likely to be accessed together are placed on the same partition.
Using a novel two-region processing approach, it then executes the \textit{hot} part of a transaction separately from the \textit{cold} part.
Our experiments show that \system{} can significantly outperform existing approaches under workloads with varying degrees of contention.
\section{Acknowledgement}
This research is supported by Google, Intel, and Microsoft as part of the MIT Data Systems and AI Lab (DSAIL) at MIT, gifts from Mellanox and Huawei, as well as NSF IIS Career Award 1453171, NSF CAREER Award \#CCF-1845763, DOE Early Career Award \#DE-SC0018947, and grant BI2011/1 of the German Research Foundation (DFG).

\newpage{}
\bibliographystyle{ACM-Reference-Format}
\bibliography{mybib} 

\end{document}